\documentclass[journal]{IEEEtran}
\usepackage{array}
\usepackage[caption=false,font=normalsize,labelfont=sf,textfont=sf]{subfig}
\usepackage{textcomp}
\usepackage{stfloats}
\usepackage{url}
\usepackage{verbatim}
\usepackage{graphicx}
\usepackage{cite}
\usepackage{cite}
\usepackage{amsmath,amssymb,amsfonts}
\usepackage{xcolor}
\usepackage{multirow}
\usepackage{enumerate}
\usepackage{bbding}
\usepackage{enumitem}

\usepackage{pifont}

\usepackage[linesnumbered,ruled]{algorithm2e}
\usepackage{algorithmicx}

\usepackage{units}
\usepackage{CJKutf8}

\begin{document}

\title{DeFT: Mitigating Data Dependencies for Flexible Communication Scheduling in Distributed Training}

\author{
    \IEEEauthorblockN{Lin Meng$^{*\dag}$, Yuzhong Sun$^{*}$\\}
    \IEEEauthorblockA{$^*$ Institute of Computing Technology, Chinese Academy of Sciences\\}
    \IEEEauthorblockA{$^\dag$ University of Chinese Academy of Sciences\\}
    \IEEEauthorblockA{\{menglin20z, yuzhongsun\}@ict.ac.cn}
}



\maketitle

\begin{abstract}
Communication scheduling aims to reduce communication bottlenecks in data parallel training (DP) by maximizing the overlap between computation and communication. However, existing schemes fall short due to three main issues: (1) hard data dependencies break some overlapping between communication and computation; (2) high coverage rates impair further improvement on performance; (3) imbalanced communication/computation times of tensors caused by partitioning/fusion strategies cause more bubbles. To address these drawbacks, we propose a new communication scheduling scheme \verb|DeFT|, whose key insight is to mitigate data dependencies and support flexible scheduling in distributed training. \verb|DeFT| uncovers new overlapping chances in training by transforming the scheduling problem into multiple knapsack problems. Specifically, \verb|DeFT| eliminates hard dependencies with delayed updates, reducing the coverage rate by adjusting update frequency and utilizing heterogeneous communication links, merging the computation times of backward or forward as the knapsack capacity to avoid the negative impact of unbalanced tensors. Additionally, \verb|DeFT| preserves training accuracy by adjusting its scheduling strategy via convergence loss quantification. Extensive experiments with 16 A100 GPUs showed that \verb|DeFT| achieved speedups of 29\% to 115\% on three representative benchmarks compared to US-Byte and Bytescheduler with no loss of accuracy. 
\end{abstract}


\section{Introduction}
\IEEEPARstart{D}{eep} Neural Networks (DNNs) have been widely used in many domains, such as Computer Vision and Natural Language Processing. Data Parallelism (DP) is a major practice for distributed training\cite{b12}. However, its performance is often far from optimum, mainly due to the communication bottleneck \cite{a9, b25}.

Previous approaches have paid attention to the communication overhead in DP. Wait-free backward propagation (WFBP) reduces communication overhead by overlapping computation and communication in backward propagation\cite{b3}. MG-WFBP \cite{b21} and other approaches \cite{b18,b50} further optimize this process through tensor fusion and have been integrated into deep learning frameworks \cite{b14}.

Building on WFBP, communication scheduling schemes \cite{a1,a2,a4,a5,a6,a7} aim to change the launching orders of communication tensors for better overlapping of communication and computation. Priority-based communication scheduling approaches \cite{a1,a2,a4,a5} preferentially transmit small tensor blocks close to the input layers in backward propagation so that the forward propagation of the next iteration can start earlier. With the strict condition that all communication overhead can be overlapped with computation, the performance of data parallelism can approach to the theoretical optimum, i.e., linear scaling \cite{b8}.

\begin{table}[t]
\centering
\setlength{\abovecaptionskip}{0.cm}
\setlength{\belowcaptionskip}{-0.cm}
\setlength{\leftskip}{-2pt}
  \caption{The computation and communication time of different DNNs}
  \resizebox{\linewidth}{7mm}{
  \begin{tabular}{ccccc}
    \hline
    DNN&$T_{forward}$&$T_{backward}$&$T_{communication}$&$CR$\\
    \hline
    ResNet-101&59ms&118ms&242ms&1.67\\
    VGG-19&37ms&93ms&258ms&1.98\\
    GPT-2&169ms&381ms&546.4ms&0.99\\
  \hline
  \end{tabular}}
\end{table}

However, existing communication scheduling approaches are still face significant challenges in achieving optimum due to the following three problems. First, some data dependencies in WFBP prevent their corresponding communications from being overlapped with computations. As shown in Fig. 1(a), the communication of bucket \#1 can only be executed after its backward computation is done in the current iteration and before the forward computation of the next iteration, which is conflicted with overlapping. Similarly, no communication can be overlapped with the backward computation of bucket \#6 and the forward computation of bucket \#6 in the next iteration, which wastes potential overlapping opportunities. We refer to these scenarios as \textbf{hard dependencies}, which may severely hinder the performances of communication scheduling schemes from achieving linear scaling.

\begin{figure*}[t]
\setlength{\abovecaptionskip}{-0.1cm}
\setlength{\belowcaptionskip}{-0.cm}
  \centering
  \includegraphics[width=\textwidth]{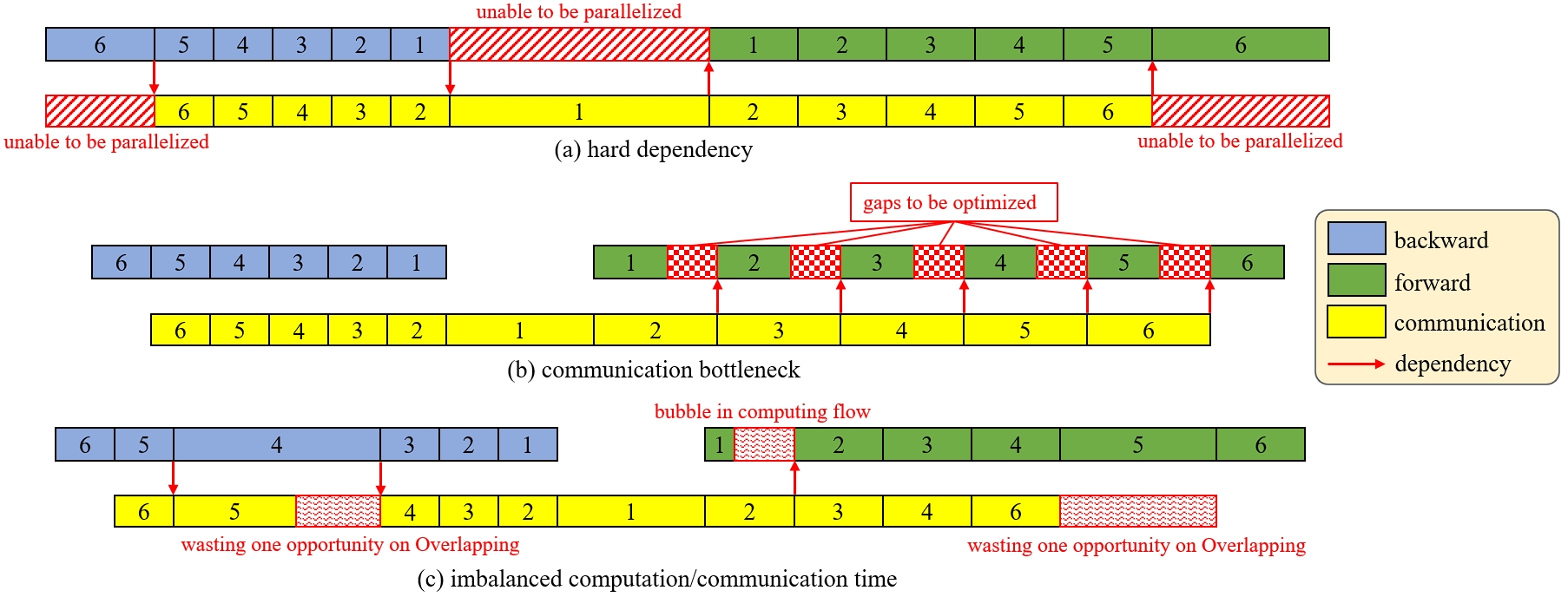}
  \caption{Three problems that cannot be solved by current communication scheduling schemes. In (a), communications/computations with hard dependencies are unable to be parallelized with the other party. In (b), communication bottlenecks cause gaps to optimal performance. In (c), imbalance in computation/communications causes wasted overlapping opportunities or bubbles.}
\end{figure*}

Secondly, although existing scheduling schemes utilize more overlapping opportunities in forward propagation stages compared to WFBP, their improvements are still constrained by the long communication overhead. Table \uppercase\expandafter{\romannumeral1} presents the forward propagation time $T_{forward}$, backward propagation time $T_{backward}$ and communication time $T_{comm}$ in one iteration of three DNNs (ResNet-101\cite{a29}, VGG-19\cite{a30} and GPT-2\cite{a31}). We define the coverage rate (CR) as $\frac{T_{comm}}{T_{forward}+T_{backward}}$. If CR is greater than 1, the excess portion of the communication can not be overlapped with computation. As shown in Fig. 1(b), the gaps are caused by bucket \#2 to \#6 when CR $>$ 1, indicating that the communication overhead of these buckets cannot be fully overlapped with computation.

Moreover, existing scheduling schemes only consider the communication time of buckets (i.e., bucket size) in their partitioning/fusion strategies, but neglect the computation time of buckets. Specifically, the computation time of each bucket depends on the computational complexity of its operators rather than the size of the bucket. In Table \uppercase\expandafter{\romannumeral2}, we list the computation times and communication times of buckets in VGG-19 with a severe imbalance. Fig. 1(c) further illustrates an example where the backward computation time of bucket \#4 is longer than the communication time of bucket \#5. Since the other buckets are not ready at the same time, the overlapping opportunity of the red part is wasted. In the next iteration, the forward computation of bucket \#1 is too short, while the remaining communication of bucket \#2 is relatively long. Such imbalances also lead to bubbles in computing stream because of the data dependency of bucket \#2's communication.

\begin{table}\small
\setlength{\abovecaptionskip}{0.cm}
\setlength{\belowcaptionskip}{-0.cm}
\setlength{\leftskip}{-2pt}
  \caption{Communication/computation times of buckets in VGG-19}
      \label{tab:vggbucket}
  \resizebox{\linewidth}{15mm}{
  \begin{tabular}{cccc}
    \hline
    Bucket id&forward(ms)&backward(ms)&communication(ms)\\
    \hline
    1&1238&72496&1968\\
   
    2&28799&12786&11262\\
    
    3&4801&4872&15447\\
    
    4&1899&2319&178643\\
    
    5&326&484&31754\\
    
    6&103&162&8651\\
    \hline
    total&37166&93119&257725\\
  \hline
\end{tabular}}
\end{table}

To address the three problems, we propose an efficient communication scheduling scheme called \verb|DeFT|. The key insight is to mitigate data \verb|De|pendencies and support \verb|F|lexible scheduling in distributed \verb|T|raining. \verb|DeFT| first introduces delayed updates, which postpone the communications associated with hard dependencies, allowing for more flexibility in scheduling. \verb|DeFT| also adaptively lowers the update frequencies and merges the communications of buckets from different iterations to reduce the coverage rates indirectly. Additionally, \verb|DeFT| introduces heterogeneous communication with multi-links, which enable concurrent communications to further reduce CR. To address the third problem, \verb|DeFT| merges the computations of all buckets into one unified overlapping capacity to temporarily neglect the data dependencies in scheduling. Finally, \verb|DeFT| utilizes the above techniques to transform the two-stage (forward and backward) communication scheduling problem into two 0/1 multi-knapsack problems.

\verb|DeFT| consists of three modules and is implemented within PyTorch \cite{b18}. The $Profiler$ module is responsible for collecting performance logs and reconstructing them at the bucket level. The $Solver$ module transforms the two-stage communication scheduling problem into two 0/1 multi-knapsack problems and solves them with a greedy algorithm. Finally, the $Preserver$ module constructs a convergence-preserving mechanism to avoid accuracy decreasing caused by reduced update frequency.

To validate the effectiveness of \verb|DeFT|, we evaluated its performance using three representative benchmarks: ResNet-101, VGG-19 and GPT-2. We compared \verb|DeFT| with popular communication scheduling schemes including PyTorch \cite{b14}, Bytescheduler \cite{a1} and US-Byte \cite{a6} on a 16-GPUs cluster with 40Gbps Ethernet bandwidth interconnection. \verb|DeFT| outperformed those schemes by up to 115\% in throughput while achieving no loss in training accuracy. The contributions of this paper are summarized as follows:
\begin{itemize}[]
    \item We propose a new communication scheduling scheme \verb|DeFT|, which can minimize the bubbles in the computing stream. The key insight is to mitigate data dependencies in DP.
    \item We transform the two-stage communication scheduling problem into two 0/1 multi-knapsack optimizations to search for the optimal scheduling. We further introduce heterogeneous links for concurrent communication to alleviate the communication bottlenecks in DP.
    \item We quantify the degree of convergence loss reduction using methods in \cite{a19} and provide an automatic adjusting mechanism for better trade-off between performance and accuracy.
    \item We implemented the prototype of \verb|DeFT| \footnote{https://github.com/meng980626/DeFT.} within PyTorch and conducted extensive experiments on clusters. Compared with state-of-the-art schemes, \verb|DeFT| significantly accelerated data parallel training with almost no loss of accuracy.
\end{itemize}

The rest of the paper is organized as follows. We introduce the background and motivation in Section \uppercase\expandafter{\romannumeral2}. In Section \uppercase\expandafter{\romannumeral3}, we provide a detailed introduction to our method, followed by our system design in Section \uppercase\expandafter{\romannumeral4}. Section \uppercase\expandafter{\romannumeral5} evaluated the performance of \verb|DeFT| compared to existing communication scheduling schemes. We present a discussion in Section \uppercase\expandafter{\romannumeral6} and introduce the related work in Section \uppercase\expandafter{\romannumeral7}. Finally, we summarize our work in Section \uppercase\expandafter{\romannumeral8}.

\section{Background}

\subsection{Data Parallelism}
Data parallelism is one of the most common methods to accelerate distributed DNN training. In data parallelism, each worker uses the same DNN parameters to perform forward and backward propagation on its own data batch in each iteration. Then, the gradients from all workers are aggregated to update the DNN parameters \cite{a8}. Since the size of gradients is large, communication in data parallelism becomes its performance bottleneck.

$\textbf{Wait-free backward propagation.}$ Several optimizations have been proposed to alleviate such communication bottlenecks. Wait-free backward propagation (WFBP) utilizes the opportunity to parallel computation and communication based on the independence of DNN’s gradient calculation \cite{b3}. Once a layer’s gradient is calculated, the communication operation is called on its entire tensor. Fig. 2 shows the process of WFBP in one iteration of DP training.

\begin{figure}[t]
\setlength{\abovecaptionskip}{-0.2cm}
\setlength{\belowcaptionskip}{-0.cm}
  \centering
  \includegraphics[width=\linewidth]{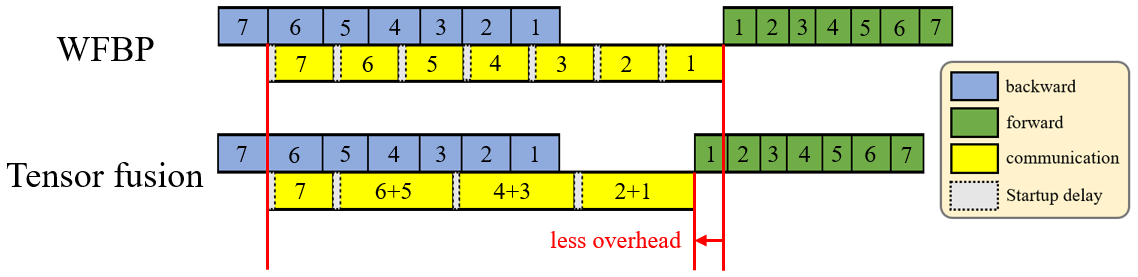}
  \caption{An example of training a 7-layer network in WFBP w/ and w/o Tensor Fusion. The communication overhead with tensor fusion is lower due to the less times of startup delays.}
\end{figure}

$\textbf{Tensor fusion.}$ Since DNNs may contain many layers, frequent communication launching in WFBP may introduce significant communication overhead. Moreover, small gradient tensor transmissions cannot fully utilize bandwidth resources. Therefore, tensor fusion optimizes WFBP by merging small layers into a large tensor called bucket to reduce the total communication overhead \cite{b14,b21}. Fig. 2 shows the improvement of tensor fusion.

$\textbf{Data Dependencies in WFBP.}$ In WFBP, computation and communication operations of buckets can be roughly divided into two streams: the computing stream and the communicating stream. Operations are executed serially within each stream but can be executed in parallel between streams. The execution of each operation is constrained by the dependencies of other operations on another stream. Fig. 2 shows examples of a dependency DAG in WFBP with tensor fusion.

\subsection{Communication Scheduling}

\begin{figure}[t]
\setlength{\abovecaptionskip}{-0.2cm}
\setlength{\belowcaptionskip}{-0.cm}
  \centering
  \includegraphics[width=\linewidth]{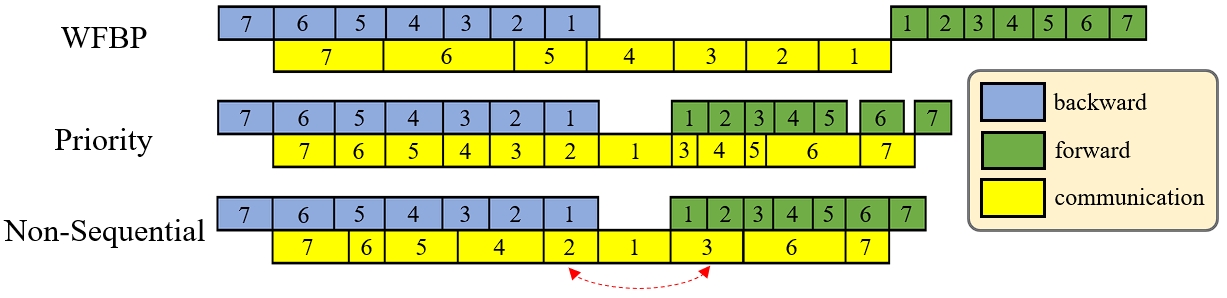}
  \caption{Difference in three communication scheduling schemes. Priority schemes utilize forward computation to increase overlapping, while non-sequential schemes have better tensor communication order and lower total communication overhead.}
\end{figure}

$\textbf{Priority scheduling.}$ The gradient tensor computation order in DNN training proceeds from the output layer to the input layer, meaning the forward propagation of the next iteration must wait until the computation and communication of the input layer are completed. However, such communication is delayed by the communications of previous layers. For example, the communication of Bucket \#1 of WFBP in Fig. 3 must wait for the completion of the previous bucket’s communication and cannot start immediately after its backward computation is completed. To address this, Bytescheduler \cite{a1} and P3 \cite{a2} propose priority scheduling, which delays part of the communications of the previous buckets through tensor partitioning to start the forward propagation of the next iteration earlier. Those methods also utilize the overlapping opportunities of forward propagation in the computing stream.

$\textbf{Tensor partition.}$ As shown in Fig. 3, tensor partition strategies first slice the original buckets into multiple tensor blocks according to partition size. The partition size is a hyper-parameter of scheduling. Priority scheduling schemes then preferentially transmit the last tensor blocks including the input layer, and the previous tensor blocks are scheduled to the forward propagation of the next iteration. Priority scheduling is also called sequential scheduling.

$\textbf{Non-sequential scheduling.}$ Although priority scheduling schemes start the forward propagation of the next iteration as early as possible, it is a sub-optimal solution for achieving the shortest iteration time. As shown in Fig. 3, for cases with varying tensor sizes, using different scheduling order can further reduce the total communication overhead. US-Byte \cite{a6} proposes a greedy algorithm with low complexity to find an approximate optimal solution for scheduling order.

\subsection{Motivation and Challenge}
Although these scheduling schemes achieve shorter iteration times, we observe that there is still significant communication overhead that cannot be overlapped with computation, resulting in bubbles in the computing stream. Additionally, some computations are not executed in parallel with communication, wasting opportunities for overlap. We summarize three reasons contributing to these situations:
\begin{itemize}[]
    \item \textbf{Hard Dependencies:} Some data dependencies prevent communication and computation from overlapping. We refer to these parts as hard dependencies.
    \item \textbf{High Coverage Rates:} The high coverage rate of data parallel tasks means the computation time available for overlapping is much smaller than the communication overhead.
    \item \textbf{Imbalance in Bucket Communication/Computation Times:} Due to the current bucket allocation method considering only bucket sizes, there may be an imbalance in computation and communication times between buckets, leading to wasted overlap opportunities.
\end{itemize}

However, several challenges exist in addressing these issues:
\begin{itemize}[]
    \item Adjusting the order of buckets cannot eliminates hard dependencies. If the overhead of those dependencies accounts for a significant proportion of the total time, current communication scheduling schemes fall far short of achieving theoretical optimal performance.
    \item The coverage rate of data parallel tasks is primarily determined by the DNN type and computing environment. While some approaches like gradient compression \cite{a10} can directly reduce communication volume, they may negatively impact training accuracy. In contrast, communication scheduling schemes aim to improve performance through better overlapping without any accuracy degradation, leading to an inability to reduce the coverage rate. Finding the optimal trade-off between training time and accuracy is challenging.
    \item The negative impact of imbalanced communication/computation time cannot be eliminated due to the strict data dependencies between communications and computations in WFBP.
\end{itemize}

In this paper, we propose \verb|DeFT| to address these three challenges. We provide a detailed introduction in Section \uppercase\expandafter{\romannumeral3} on how \verb|DeFT| integrates these three solutions and transforms communication scheduling into solving a 0/1 knapsack problem. In Section \uppercase\expandafter{\romannumeral4}, we introduce other implementation details of \verb|DeFT|, including the algorithm’s lifecycle during training, the implementation of the $Profiler$ module and the mechanism for preserving accuracy in the $Preserver$ module. Table \uppercase\expandafter{\romannumeral3} presents a comparison of DeFT with several other communication scheduling schemes.

\begin{table*}[t]
\setlength{\abovecaptionskip}{0.cm}
\setlength{\belowcaptionskip}{-0.cm}
\setlength{\leftskip}{-2pt}
  \caption{Comparison of Four Scheduling Schemes}
  \label{tab:comparison}
  \resizebox{\textwidth}{18mm}{
  \begin{tabular}{p{13mm}p{9mm}p{18mm}p{25mm}p{15mm}p{2cm}p{2cm}}
    \hline
    scheme&forward overlap&tensor fusion&scheduling strategy&convergence consistency&hard dependency&performance\\
    \hline
    PyTorch& \ding{55} & regular \& uniform & - & baseline & exist & limited by CR\\
    Bytescheduler& \checkmark & auto-tune \& uniform & Sequential & \checkmark & exist& limited by CR\\
    US-Byte& \checkmark & unequal-sized & Non-Sequential & \checkmark & exist& limited by CR\\
    DeFT& \checkmark & unequal-sized (constrained) & 0/1 Multi Knapsack& approximate & eliminate by delayed update& adaptively reduce update frequency\\
  \hline
\end{tabular}}
\end{table*}

\section{Method}

\subsection{Overview}
As mentioned in Section \uppercase\expandafter{\romannumeral1} and \uppercase\expandafter{\romannumeral2}.C, although priority scheduling and non-sequential scheduling schemes minimize iteration time by adjusting the order of bucket communications, they still fail to achieve the theoretical optimal performance of DP due to (1) hard dependencies, (2) communication bottlenecks, and (3) imbalanced bucket computation/communication time.

To address the first problem, \verb|DeFT| modifies the strategy for updating parameters. When there exists hard dependencies before and after a communication, \verb|DeFT| delays that communication to eliminate the corresponding bubbles. However, this way results in the gradients of these buckets not being synchronized. Therefore, \verb|DeFT| accordingly delays the timing of parameter updates if all gradients of one iteration are not fully synchronized.

To address the second problem of communication bottlenecks, \verb|DeFT| adaptively lowers the frequencies of parameter updates to reduce the total communication volume. Specifically, the goal of \verb|DeFT| is to reduce the communication overhead in each iteration to be less than the computation time (including forward and backward) so that the communication can be covered by computation. To achieve this, \verb|DeFT| delays the communications of some buckets selected by our algorithm in each iteration and stores them locally. When those locally accumulated buckets contain all buckets of a complete old iteration, \verb|DeFT| merges them with buckets of current iteration to reduce the total communication volume. Assuming the coverage ratio is N:M (N$\geq$M), the update frequency of \verb|DeFT| will be approximately M times in N iterations. We will introduce more details of our algorithm in Section \uppercase\expandafter{\romannumeral3}.B.

For the third problem, we believe that communications should be scheduled with more flexibility. The key insight is to mitigate data dependencies in WFBP. Therefore, \verb|DeFT| merges those imbalanced bucket computation times together and view them as one whole scheduling capacity, so that we can accommodate communications without restrictions of data dependencies.

Based on the above three techniques, \verb|DeFT| transforms the communication scheduling problem into the $0/1$ $knapsack$ $problem$ to maximizes the overlap between computation and communication. Specifically, \verb|DeFT| solves the 0/1 knapsack problem in the forward and backward stages, using computing time as the capacity of the $knapsack$ and the communication time of each bucket as the weight and profit of each $item$. The buckets delayed due to hard dependencies will be stored locally and may be put into a $knapsack$ in subsequent iterations. When there are too many delayed buckets, \verb|DeFT| merges the gradient buckets of current iteration with the locally accumulated gradient buckets of past iterations, which reduces the update frequency of the corresponding steps (similar to gradient accumulation). We will introduce more details in Section \uppercase\expandafter{\romannumeral3}.B.

\subsection{Problem formulation and Solution}
\verb|DeFT| models communication scheduling as a 0/1 knapsack problem. The problem can be modeled simply as follows:

{
\setlength{\parindent}{0cm}
$\mathbf{Problem}$ $\mathbf{1.}$ $Given$ $N$ $buckets$ $where$ $each$ $bucket$ $has$ $its$ $communication$ $time$ $and$ $its$ $computation$ $time$ $(either$ $forward$ $or$ $backward).$ $The$ $knapsack$ $capacity$ $is$ $the$ $sum$ $of$ $the$ $computation$ $times$ $of$ $all$ $buckets.$ $The$ $task$ $is$ $to$ $put$ $the$ $communications$ $of$ $the$ $buckets$ $into$ $the$ $knapsack$ $such$ $that$ $the$ $sum$ $of$ $the$ $communication$ $time$ $is$ $the$ $maximum$ $possible.$
$The$ $naive$ $mathematical$ $model$ $is$ $as$ $follows:$ 
\begin{displaymath}
    Maximize\sum_{i=1}^{n} c_i x_i
\end{displaymath}
\begin{displaymath}
    Subject\enspace  to
    \begin{cases}
        \sum_{i=1}^{n} c_i x_i \leq \sum_{i=1}^{n} t_i & \\
        x_i \in \{0,1\}, 1 \leq i \leq n
    \end{cases}
\end{displaymath}

$where$ $c_i$ $is$ $the$ $communication$ $time$ $of$ $bucket$ $i,$ $t_i$ $is$ $the$ $computation$ $time$ $of$ $bucket$ $i.$
}

\begin{figure*}[t]
\setlength{\abovecaptionskip}{-0.1cm}
  \centering
  \includegraphics[width=\textwidth]{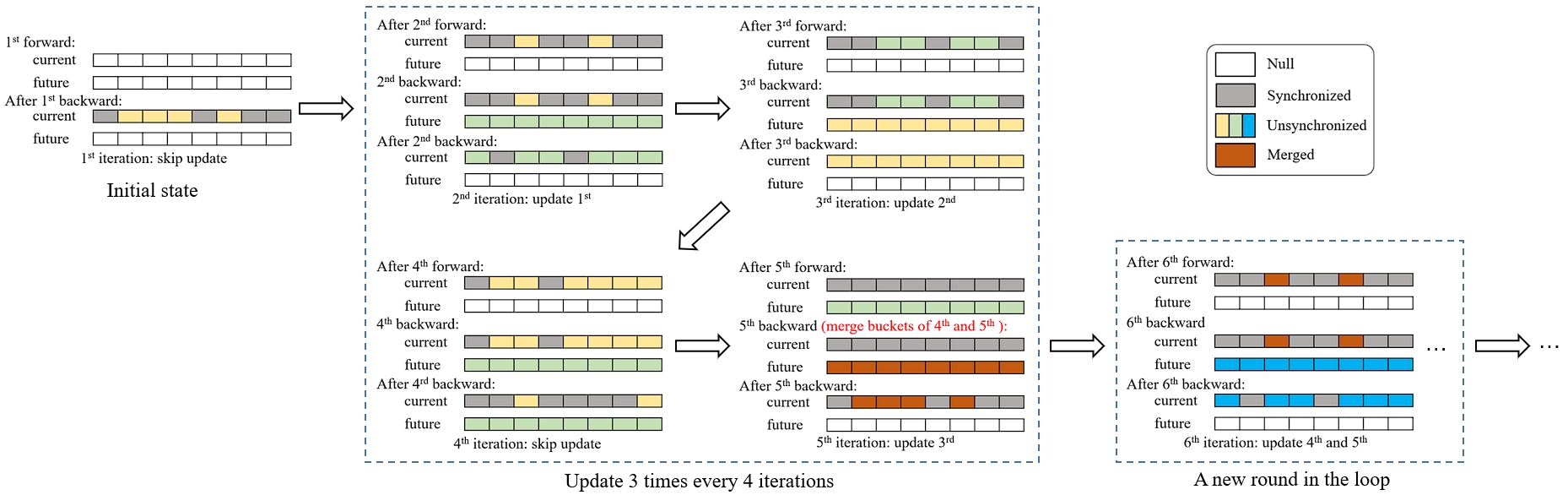}
  \caption{The process of how current task and future task queue change. Buckets with yellow, green and blue colors represents unsynchronized gradient buckets from different iterations. In fifth iteration, green buckets in future task queue are merged with the new buckets of the fifth iteration (not shown in figure).}
\end{figure*}

\begin{algorithm}[t]
    \SetAlgoLined 
	\caption{RecursiveKnapsack}
	\KwIn{$CommTimeList=\{C_N,C_{N-1},...,C_1\},$\\ 
                $\hspace*{3.05 em} remainTime$}
        \KwOut{$ScheduleOrder=\{A_1,A_2,...,A_M\},\ where$\\$\hspace*{3.65 em} M\leq N,A_i\in[1,N],\forall i,j \in [1,M], A_i\neq A_j$\\ }
        \If{$CommTimeList\ is\ \emptyset$}{
            \textbf{return} $\emptyset$\\
        }
	$order1=NaiveKnapsack(CommTimeList,remainTime)$\;    
        $order2=RecursiveKnapsack(CommTimeList-C_N,remainTime-T_{N-1})$\;
        \eIf{$sum(order1)>sum(order2)$}{
            \textbf{return} $order1$
        }{
            \textbf{return} $order2$
        }
\end{algorithm}

The solution of \verb|DeFT| is to delay the launching of parameter updates when only a subset of buckets generated in one iteration is synchronized. Assuming that \verb|DeFT| schedules $b$ and $f$ buckets on average at the backward and forward stages of each iteration respectively, and satisfies $b+f$\textless$n$, where $n$ is the total number of buckets. Hence, \verb|DeFT| reduces communication operations by $n-(b+f)$ buckets per iteration compared to original training, such that it will only update $b+f$ times in $n$ iterations. The coverage rate can be approximated to $\nicefrac{n}{(b+f)}$.

For implementation, \verb|DeFT| establishes two queues, called the current task queue and the future task queue. The current task queue is used to store the remaining buckets from previous iterations. If there are already some buckets in the current task queue while new buckets of current iteration are arrived, these new buckets will be stored in the future task queue to avoid confusion with the current task queue. The buckets in the future task queue will be copied to the current task queue once the current task queue is cleared (i.e., all buckets in the current task queue are synchronized and updated to parameters). The future task queue always retains all buckets' results from one or more iterations. If the next iteration’s backward stage starts and there are still buckets in the future task queue, \verb|DeFT| will merge the two results together, so that these two results only need to be synchronized once to reduce the total communication volume.

Specifically, there are four cases to handle two queues:
{
\setlength{\parindent}{0cm}
$\mathbf{Case}$ $\mathbf{1.}$ In the forward stage, \verb|DeFT| needs to schedule the remaining buckets in the current task queue. Since there are no dependencies between buckets in the current task queue and forward computations, \verb|DeFT| simply solves a naive 0/1 knapsack optimization, where the item list is the current task queue and the knapsack capacity is the sum of forward computation times.

$\mathbf{Case}$ $\mathbf{2.}$ At the beginning of the backward stage, if there are buckets in the current task queue that have not been synchronized before, \verb|DeFT| first estimates whether the total backward computation time can cover the communications of all remaining buckets. If it is not enough, \verb|DeFT| uses these buckets as the item list and the total backward computation time as the knapsack capacity to solve a naive 0/1 knapsack problem, since these old buckets have already been calculated in the previous iterations and have no dependencies with current backward computations. The new buckets of this iteration are stored (or merged with previous buckets) in the future task queue locally.

$\mathbf{Case}$ $\mathbf{3.}$ Otherwise, if the total backward computation time is long enough for communications of all remaining buckets in the current task queue, \verb|DeFT| preferentially schedules the communications of these remaining buckets, while the new gradient buckets in this iteration will be stored (or merged) into the future task queue temporarily. Afterwards, \verb|DeFT| uses the future task queue as the $itemlist$ and the remaining time as the knapsack capacity to solve the new 0/1 knapsack problem with recursion (Algorithm 1). Finally, \verb|DeFT| copies the remaining buckets in the future task queue into the current task queue. Additionally, one parameter update will be performed after this backward stage since all buckets in the last current task queue have already been synchronized.
}

{
\setlength{\parindent}{0cm}
$\mathbf{Case}$ $\mathbf{4.}$ If all buckets in the current task queue are already completed before the backward stage, \verb|DeFT| directly uses Algorithm 1 to recursively select the bucket for communication in this backward stage. If the future task queue is not empty, \verb|DeFT| merges the buckets in the future task queue to the buckets generated by current iteration before its communication. Finally, similar to Case 3, \verb|DeFT| copies the remaining buckets from the future task queue to the current task queue and performs a parameter update at the end of this iteration.
}

Algorithm 2 shows the pseudo-code of \verb|DeFT|’s complete process for solving the scheduling scheme in two stages. Figure 4 presents an example of the transition processes of the current task and future task queues respectively. In Fig. 4, the initial state is a special case of Case 4 when there are no buckets in the current task queue at the beginning of the backward stage. Then, \verb|DeFT| utilizes Algorithm 1 to solve the backward stage. Then, \verb|DeFT| utilizes Algorithm 1 to solve the problem via recursion. In the second iteration’s forward stage, the situation corresponds to Case 1, and \verb|DeFT| solves it as a naive 0/1 knapsack problem. In the second iteration’s backward stage, the situation corresponds to Case 3, and \verb|DeFT| first transmits the remaining two yellow buckets from the first iteration in DP, then uses Algorithm 1 to select buckets from

\begin{figure}[hb]
  \centering
  \includegraphics[width=\linewidth]{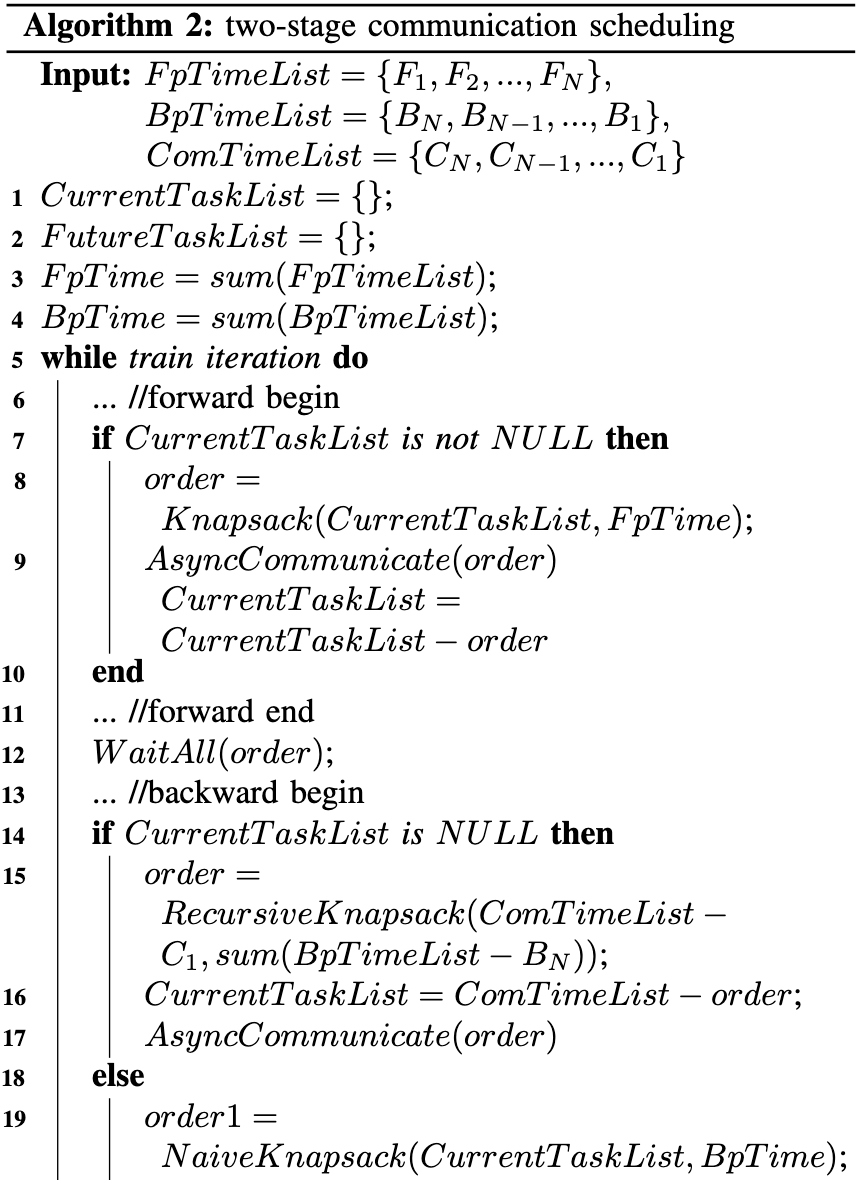}
\end{figure}

{
\setlength{\parindent}{0cm}
the eight green ones generated in this iteration. At the end of this iteration, \verb|DeFT| updates the parameters with the gradient buckets from the first iteration since all these yellow buckets have already been synchronized. In the fourth iteration’s backward stage, the situation corresponds to Case 2. \verb|DeFT| selects buckets in the current task queue using the naive 0/1 knapsack solution, and the new eight green buckets can only be stored in the future task queue since there are still two remaining yellow buckets in the current task queue. In the backward stage of the fifth iteration, the situation corresponds to Case 4, and the green buckets from the fourth iteration in the future task queue are merged with the new buckets from the fifth iteration. After all red buckets have been synchronized, these gradients from the fourth and fifth iterations will be updated together at the end of the sixth iteration.
}

\begin{figure}[t]
  \centering
  \includegraphics[width=\linewidth]{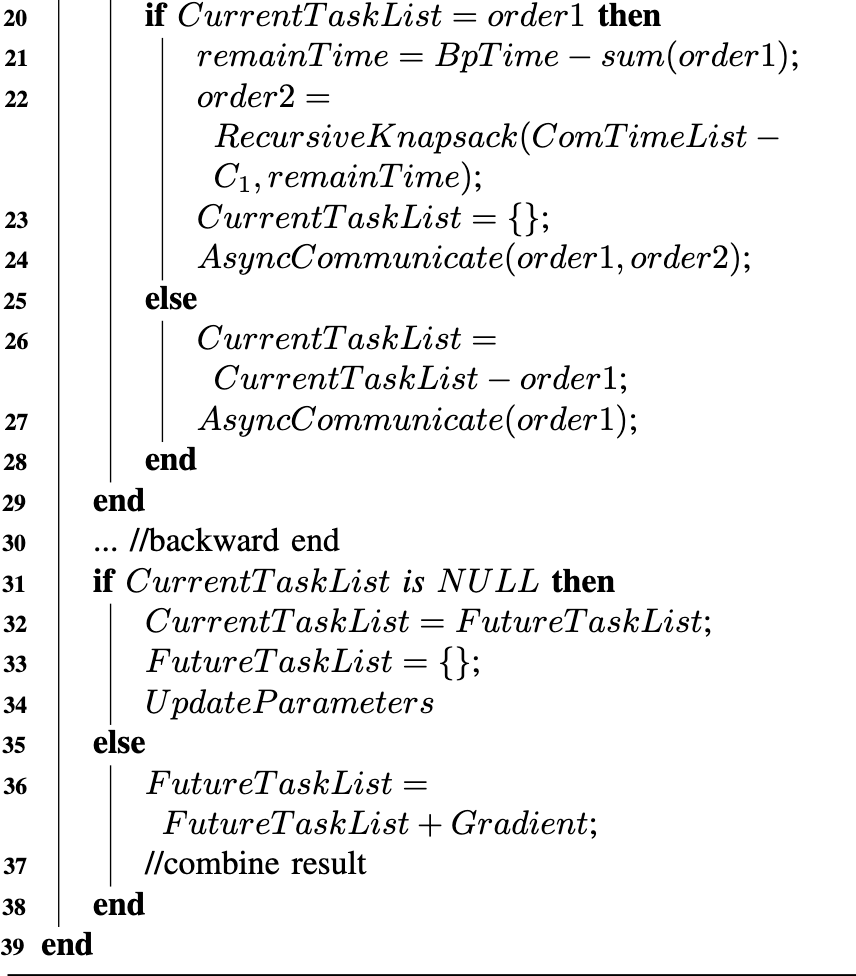}
\end{figure}

\subsection{Heterogeneous communication}
We further introduce heterogeneous communication, which can increase the update frequency of \verb|DeFT| closer to the baseline to avoid accuracy decreasing (The detailed discussion of training accuracy will be given in Section \uppercase\expandafter{\romannumeral4}.C). The key idea is to concurrently communicate a portion of buckets via heterogeneous links (e.g., CPU and GPU).

Previous work \cite{a12} has already improved communication efficiency by leveraging concurrent communication. In this work, we utilize heterogeneous links to achieve concurrent communication using different communication libraries. Specifically, a small portion of buckets communicate through the gloo \cite{a14} communication library, while other buckets use the NCCL \cite{a15} communication library. After introducing heterogeneous communication, an example of \verb|DeFT|’s scheduling result is shown in Fig. 5. The communication of bucket \#7 can be paralleled with the communication of buckets \#6, \#5, \#3, and \#2.

\begin{figure}[t]
\setlength{\abovecaptionskip}{-0.2cm}
\setlength{\belowcaptionskip}{-0.cm}
  \centering
  \includegraphics[width=\linewidth]{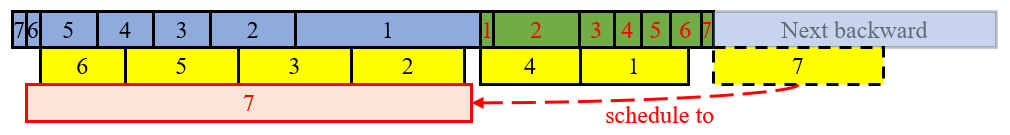}
  \caption{An example of concurrent heterogeneous communication. The communication of bucket \#7 is scheduled to the heterogeneous link in the backward of last iteration.}
\end{figure}

\begin{figure}[t]
\setlength{\abovecaptionskip}{-0.2cm}
\setlength{\belowcaptionskip}{-0.cm}
  \centering
  \includegraphics[width=\linewidth]{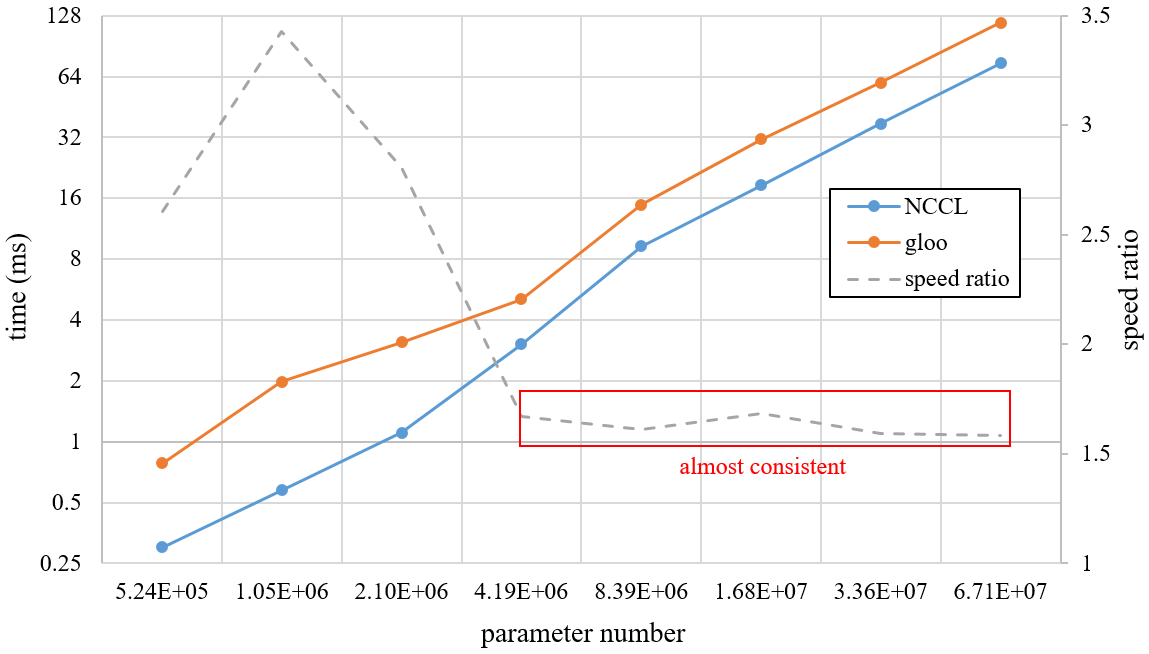}
  \caption{The communication time of all-reduce along with the size of parameters with datatype float32 using different communication libraries. When the number of parameters is greater than 4000000, the speed ratio of the two communication libraries remains essentially constant.}
\end{figure}

Due to the slower communication speed with gloo \cite{a13}, the communication of bucket on gloo is longer than it on NCCL. We first measured the communication speed difference for different bucket sizes on the two communication libraries. As Fig. 6 shows, the ratio in communication speed between the two heterogeneous communication libraries is almost consistent when the bucket size exceeds a certain value. Specifically, NCCL is 1.59 to 1.69 times faster than gloo when the bucket size is more than 4,194,304. Since the default bucket size is 25MB (6553600 fp32 parameters) in PyTorch, \verb|DeFT| increased the number of knapsacks in $\mathbf{Problem}$ $\mathbf{1}$ to two, corresponding to NCCL communication and gloo communication, respectively. The capacity of the NCCL communication knapsack is $\mu$ times that of the gloo communication, where $\mu$ is a constant number representing the speed ratio between the two communication libraries. In our experiments, $\mu$ is set to 1.65.

Furthermore, we mitigate interference and contention in communications by designating distinct network interface cards (NICs) for different communication libraries within the PyTorch framework. Specifically, each node is equipped with multiple NICs, and the PyTorch framework is capable of assigning different NIC to different communication libraries via the environment variables \verb|NCCL_SOCKET_IFNAME| and \verb|GLOO_SOCKET_IFNAME|. We evaluated the concurrent communication performance of two communication libraries, NCCL and gloo, in both multi-link (using different NICs for 2 communication libraries) and single-link (using the same NIC for 2 communication libraries) modes. Table \uppercase\expandafter{\romannumeral4} presents a comparative analysis. NCCL's communication speed remains consistent in both modes. In contrast, gloo's performance is consistent when the tensor size is small, but it exhibits a significant speedup of around 20\% in multi-link mode compared to single-link mode when the tensor size is large. This suggests that resource contention occurs in that case, and using multi-link with multi-library approach can mitigate the performance degradation caused by resource contention.

\begin{table}[t]
\setlength{\abovecaptionskip}{0.cm}
\setlength{\belowcaptionskip}{-0.cm}
\setlength{\leftskip}{-2pt}
  \caption{The All-Reduce time of multi-link and single-link.}
  \label{tab:multi-link}
  \resizebox{\linewidth}{7mm}{
  \begin{tabular}{ccccccc}
    \hline
    \multicolumn{2}{c}{Tensor size}&4194304&8388608&16777216&33554432&67108864\\
    \hline
    \multirow{2}{*}{multi-link} &gloo&22ms(+0\%)&\textbf{41ms(+18\%)}&\textbf{80ms(+17\%)}&\textbf{169ms(+17\%)}&\textbf{428ms(+20\%)}\\
                                &NCCL&14ms&25ms&51ms&110ms&231ms\\
    \multirow{2}{*}{single-link}&gloo&22ms&\textbf{50ms}&\textbf{96ms}&\textbf{204ms}&\textbf{534ms}\\
                                &NCCL&13ms&26ms&53ms&110ms&230ms\\
  \hline
\end{tabular}}
\end{table}

After introducing heterogeneous communication, we transformed $\mathbf{Problem}$ $\mathbf{1}$ (a 0/1 knapsack problem) into $\mathbf{Problem}$ $\mathbf{2}$ (a 0/1 multi-knapsack problem):

{
\setlength{\parindent}{0cm}
$\mathbf{Problem}$ $\mathbf{2.}$ $Given$ $N$ $buckets$ $where$ $bucket$ $j$ $has$ $communication$ $time$ $c_j$ $and$ $computation$ $time$ $t_j$ $(forward$ $or$ $backward).$ $There$ $are$ $2$ $knapsacks,$ $one$ $with$ $a$ $capacity$ $equal$ $to$ $the$ $sum$ $of$ $computation$ $times$ $of$ $all$ $buckets$ $and$ $the$ $other$ $with$ $a$ $capacity$ $equal$ $to$ $\mu$ $times$ $that$ $of$ $the$ $first$ $knapsack$ $(\mu$ $is$ $the$ $speed$ $ratio$ $of$ $gloo$ $and$ $NCCL$ $communication).$ $The$ $task$ $is$ $to$ $put$ $the$ $communications$ $of$ $buckets$ $into$ $the$ $2$ $knapsack$ $such$ $that$ $the$ $sum$ $of$ $the$ $communication$ $time$ $associated$ $with$ $them$ $is$ $the$ $maximum$ $possible.$
$The$ $naive$ $mathematical$ $model$ $is$ $as$ $follows:$ 
\begin{displaymath}
    Maximize\sum_{i=1}^{2}\sum_{j=1}^{N} c_j x_{ij}
\end{displaymath}

\begin{displaymath}
    Subject\enspace  to
    \begin{cases}
        \sum_{j=1}^{N} c_j x_{1j} \leq \sum_{j=1}^{N} t_j & \\
        \sum_{j=1}^{N} c_j x_{2j} \leq \mu *\sum_{j=1}^{N} t_j & \\
        \sum_{i=1}^{2} x_{ij} \leq 1 & \\
         x_{ij} \in \{0,1\}, 1 \leq i \leq 2, 1 \leq j \leq N
    \end{cases}
\end{displaymath}
}

The 0/1 multi-knapsack problem is a NP-hard problem \cite{a16}. We use a greedy strategy heuristic to solve this problem, which has the advantage of low cost. Specifically, we first sort the capacity of each knapsack and the time of each bucket, and then start with the backpack with smaller capacity, and try to prioritize placing the bucket with longer time. The time complexity of the placement phase is O(N*M), where N is the number of items (buckets) and M is the number of knapsacks (communication links). Although the algorithmic complexity of solving multiple knapsacks is relatively high than single knapsack, our problem satisfies the following conditions: (1) \verb|DeFT| only have two knapsacks; (2) the knapsack capacity is not much larger than the item size; and (3) the number of items is not large (less than 20), so the overhead of solving Problem 2 is not very high. In all experiments we conducted, the overheads were always less than 1 seconds. Compared to hours of training, such overheads are acceptable.

\begin{figure}[t]
\setlength{\abovecaptionskip}{-0.1cm}
\setlength{\belowcaptionskip}{-0.cm}
  \centering
  \includegraphics[width=\linewidth]{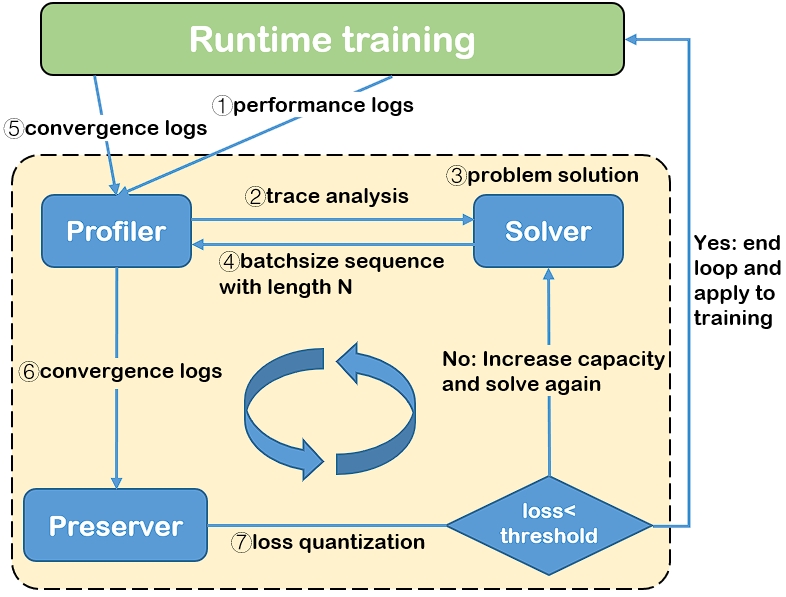}
  \caption{The system overview of DeFT.}
\end{figure}

\subsection{Tensor partition}

\verb|DeFT|’s scheduling strategy is based on solving the 0/1 knapsack problem. However, the default bucket allocation strategy may causes buckets too large with extended communication times \cite{a11}, which makes it challenging to fit these large buckets as items into knapsacks. As a solution, we have adopted the partition strategies of Bytescheduler and US-byte with additional constraints imposed. Initially, buckets are partitioned based on a partition size parameter (default to 6,500,000). Subsequently, the partition strategy of US-byte is employed to re-partition with variable partition sizes aimed at reducing the total communication overhead. Ultimately, \verb|DeFT| ensures that the communication time of the largest bucket remains less than the smallest knapsack capacity (typically the forward time divided by $\mu$), and re-partitions any bucket that does not meet this constraint.

\section{Implementation}

\subsection{Overview}
We implemented DeFT within the PyTorch framework. As shown in Fig. 7, \verb|DeFT| contains three modules: $Profiler$, $Solver$, and $Preserver$. During the early stage of training, the $Profiler$ first collects performance logs and converts them from the operator-level to the bucket-level through trace analysis tool. Subsequently, it computes the buckets’ computation and communication times and transfer them to the $Solver$, which outputs the scheduling orders using the method described in Section \uppercase\expandafter{\romannumeral3}. \verb|DeFT| temporarily applies that scheduling to the training process for several iterations. Simultaneously, the $Profiler$ collects convergence-related information and submits it to the $Preserver$. The $Preserver$ first approximates that scheduling as a variable batch size sequence. Then , it quantifies the loss difference between that sequence and the original training using the method in \cite{a19}. If the loss difference is less than an empirical threshold, it indicates that the scheduling result has almost no effect on convergence, which prompts \verb|DeFT| to apply that scheduling result to the training process. However, if the loss difference exceeds the threshold, \verb|DeFT| increases the knapsack capacity in Problem 2 and resolves it. Increasing the knapsack capacity allows for more communications in each iteration, which avoids excessive decrease in parameter update frequency to preserve accuracy.

\begin{figure*}[t]
\setlength{\abovecaptionskip}{-0.1cm}
\setlength{\belowcaptionskip}{-0.cm}
  \centering
  \includegraphics[width=\textwidth]{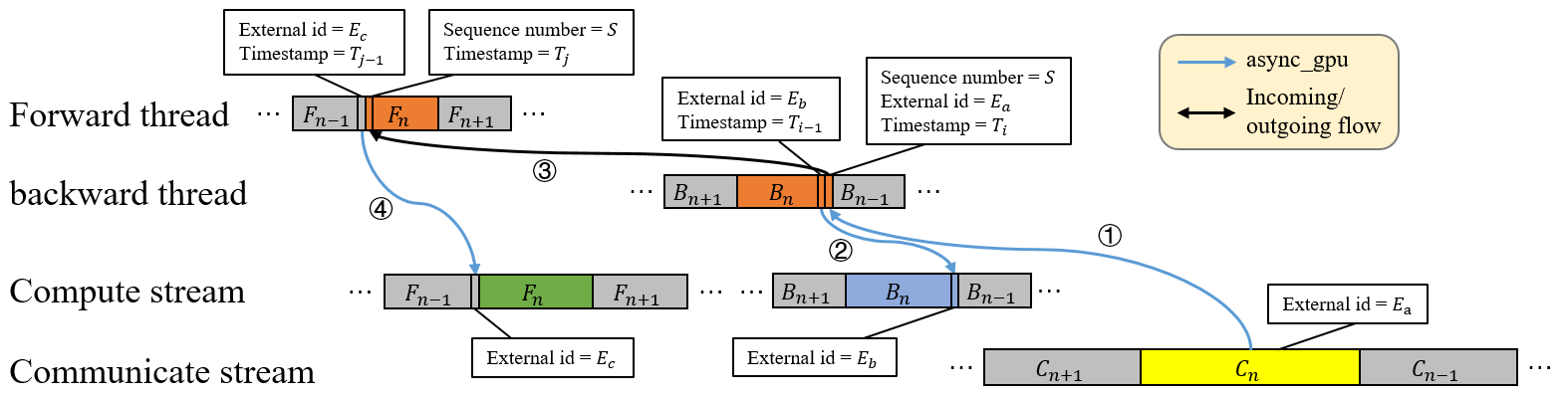}
  \caption{The 4-step analysing process of the raw logs by $Profiler$ of DeFT.}
\end{figure*}

\subsection{Trace analysis of $Profiler$}

The $Profiler$ module of \verb|DeFT| first utilizes performance analysis tools (i.e., NVIDIA Nsight Systems \cite{a20}) to obtain raw logs of training operators, including information such as kernel name, thread ID, timestamp, etc. However, since the execution time of an operator is typically milliseconds-level or even microseconds-level, the raw logs are too fine-grained for communication scheduling. Therefore, we reconstruct the raw logs from the $operator$-level to the $bucket$-level by analyzing the dependencies between operators. The analysis process is illustrated in Fig. 8.

The $Profiler$ first identifies the External ID of communication operators, since each communication operator has a one-to-one correspondence to a bucket. Using the External ID of communication operator of Bucket \#N, the $Profiler$ finds the last operator of Bucket \#N in the backward thread. Secondly, the $Profiler$ finds the penultimate operator of Bucket \#N in the backward thread through the timestamp of that last operator. The penultimate operator will launch the kernel and correspond to the last operator of Bucket \#N in the computing stream, allowing the $Profiler$ to identify the ending point of Bucket \#N in the computing stream. Next, the $Profiler$ finds the corresponding operator (i.e., the first operator of Bucket \#N) in the forward thread of the last operator in the backward thread. The $Profiler$ locates the last operator of the forward computation for Bucket \#N-1 by using the timestamp of the first operator in Bucket \#N. It then identifies the endpoint of Bucket \#N-1 based on the External ID of this operator.

After repeating this process for other buckets, the $Profiler$ obtains each bucket’s forward/backward ending point in the computing stream to calculate the computation and communication time for each bucket.

\subsection{Accuracy preserving mechanism of $Preserver$}

In this section, we first approximate the scheduling of \verb|DeFT| into training with a variable batch size sequence (loop every N iterations) to understand its convergence more easily. Subsequently, we employ the method in \cite{a19} for quantifying the convergence difference between that variable batch size sequence and the fixed original batch size. Furthermore, we implement an automatic adjustment scheme when that difference is too large and has negative impact on convergence. Such strategy helps \verb|DeFT| to make a better trade-off between performance and convergence.

\subsubsection{Variable batch size}
We first explain why \verb|DeFT|'s convergence is approximate to training with a looped and variable batch size sequence. Gradient accumulation \cite{a21} is a method for simulating large batch sizes, which requires smaller amount of memory compared to using large batch sizes directly. Typically, the original training uses an optimizer to update the model parameters after gradient calculation is completed in each iteration. Instead of updating parameters at the end of each iteration, gradient accumulation stores gradients locally without updating parameters and continuously accumulates the gradient results for N iterations before updating parameters. For convergence, gradient accumulation is equivalent to enlarging the original batch size by N times. Fig. 9 shows the difference between gradient accumulation and original training.

\begin{figure}[t]
\setlength{\abovecaptionskip}{0.cm}
\setlength{\belowcaptionskip}{-0.cm}
  \centering
  \includegraphics[width=\linewidth]{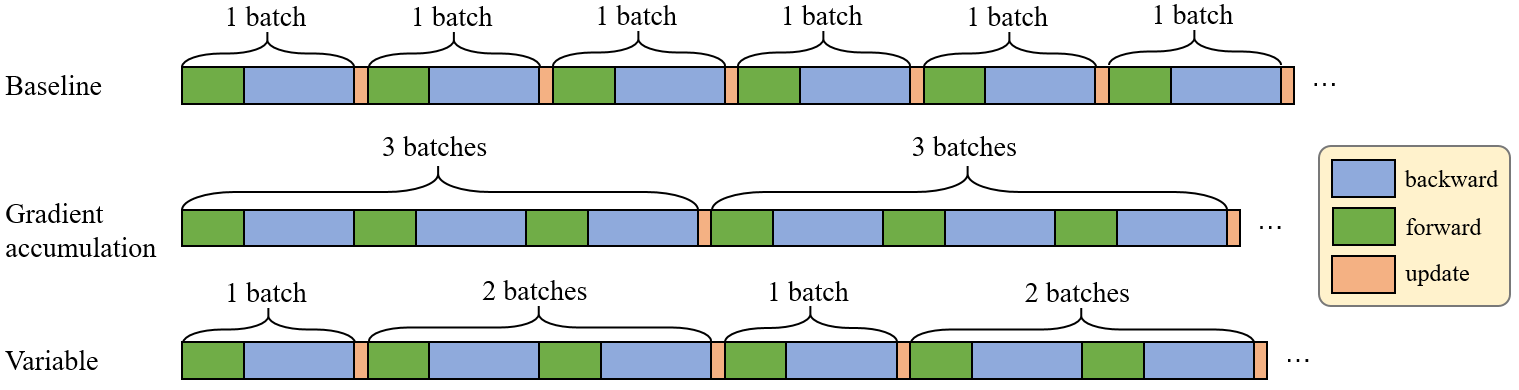}
  \caption{Different update frequencies between original DP with fixed batch size, Gradient accumulation and DeFT with variable batch size.}
\end{figure}

In \verb|DeFT|, due to the delayed update timing, the new gradient buckets sometimes merge with the previous gradient buckets (as seen in Case 2 and Case 4 in Section \uppercase\expandafter{\romannumeral3}). This merging is similar to gradient accumulation because both of them accumulated their gradient locally before synchronizations. Therefore, those merging iterations in \verb|DeFT| can be temporarily enlarging the batch size. Fig. 9 shows an example of the update timing in \verb|DeFT|. The gradient of the first batch is updated immediately after the first iteration, while the gradients of the next two batches are merged together and updated after the third iteration. Afterwards, \verb|DeFT| updates with that variable batch size sequence and loops every 3 iterations. The batch size sequence of \verb|DeFT| is defined as follows:

$Suppose$ $the$ $batch$ $size$ $of$ $the$ $original$ $training$ $is$ $B.$ $As$ $mentioned$ $above,$ $\verb|DeFT|'s$ $scheduling$ $strategy$ $can$ $be$ $regarded$ $as$ $updating$ $with$ $a$ $variable$ $batch$ $size$ $sequence$ $every$ $N$ $iterations.$ $The$ $sequence$ $of$ $batch$ $size$ $is$ $K={k_1,k_2,...,k_m},$ $satisfies,$

\begin{displaymath}
    \sum_{i=1}^{m} k_i = N, k_i \in \mathbb{N}^+
\end{displaymath}

{
\setlength{\parindent}{0cm}
$which$ $means$ $\verb|DeFT|$ $updates$ $parameters$ $in$ $batch$ $size$ $sequence$ $of$ $k_1B,k_2B,...,k_mB$ $every$ $N$ $iterations$ $(designated$ $O_D),$ $while$ $the$ $sequence$ $of$ $original$ $training$ $with$ $fixed$ $batch$ $size$ $in$ $N$ $iterations$ $is$ $B,B,...,B$ $including$ $N$ $instances$ $of$ $B$ $(designated$ $O_B).$
}

\subsubsection{Loss quantification}
Based on the above transformation, we further provide a method for quantifying the convergence of \verb|DeFT|. Specifically, we employ the method described in \cite{a19} to measure the convergence difference between \verb|DeFT| and baseline. This analysis allows us to understand how \verb|DeFT|'s scheduling strategy affects the training convergence, followed by ensuring that the accuracy remains within acceptable range.

Yin et al. \cite{a19} discussed how to dynamically determine batch size during the training process and quantified their method’s impact on convergence when using variable batch size updates. Specifically, they modeled the estimation of one batch gradient as the sum of the real gradient plus the “noise variance” introduced by each data point. Subsequently, the learning process is modeled as a random walk game with a Gaussian distribution, where the convergence efficiency of using variable batch sizes can be quantified. Drawing inspiration from this method, we provide an approach to quantify the impact of \verb|DeFT|’s scheduling scheme on convergence.

Suppose the state of the random walk game is $s_t$, i.e., the training loss at the $t_{th}$ iteration. For $s_{t+1}$, there are two possible paths of state transfer, corresponding to two situations during the learning process. The first path is directly decreasing and approaching the objective value (the lowest point of training loss), and the second path is rebounding to a higher point after reaching the target value. $s_{t+1}$ can be represented as:

\begin{displaymath}
    s_{t+1}=
    \begin{cases}
        s_t-\eta\Delta s_t & ,if\ s_t-\eta\Delta s_t \geq S^* \\
        2S^*+\eta\Delta s_t-s_t & ,otherwise\\
    \end{cases}
\end{displaymath}

where $\eta$ is the learning rate, $S^*$ is the objective value. $\Delta s_t$ is the Gaussian dice for generating moving steps and satisfies Gaussian distribution $\Delta s_t \sim \mathcal{N}(\mu_t,\frac{\sigma_t^2}{B})$. $\mu_t$ is the square sum of the gradient in iteration $t$ and $B$ is the batch size, while $\sigma_t$ is its multiplication with covariance matrix.

\begin{table}
\setlength{\abovecaptionskip}{0.cm}
\setlength{\belowcaptionskip}{-0.cm}
\setlength{\leftskip}{-2pt}
  \caption{$E_B^{s_t}(s_{t+1})$ of $O_B$ and $O_D$ when training ResNet-101}
  \resizebox{\linewidth}{7mm}{
  \begin{tabular}{ccc|cc|cc|cc|cc|c}
    \hline
    Setting&\multicolumn{11}{c}{A = 1000, N = 4, $S^*$ = 0, $\eta$ = 0.01}\\
    \hline
    {}&\multicolumn{2}{c}{iteration A}&\multicolumn{2}{c}{iteration A+1}&\multicolumn{2}{c}{iteration A+2}&\multicolumn{2}{c}{iteration A+3}&\multicolumn{2}{c}{iteration A+4}&ratio\\
    {}&B&$E_B$&B&$E_B$&B&$E_B$&B&$E_B$&B&$E_B$&\multirow{3}{*}{0.993}\\
    $O_B$&256&0.2103&256&0.2054&256&0.1989&256&0.1967&256&0.1922&\\
    $O_D$&256&0.2103&512&0.2012&\multicolumn{2}{c|}{-}&256&0.1979&256&0.1935&\\
  \hline
\end{tabular}}
\end{table}

The expected value of next state can be expressed as:
\begin{displaymath}
\begin{aligned}
  E_B^{s_t}(s_{t+1})=&(s_t-S^*-\eta \Delta s_t)\{\Phi(a)-\Phi(-a)\}+\\&
  \frac{\eta\sigma_t}{\sqrt{B}}\sqrt{\frac{2}{\pi}}e^{-\frac{a^2}{2}}+S^*
\end{aligned}
\end{displaymath}

\begin{displaymath}
    where\ a=\frac{s_t-S^*-\eta \mu_t}{\eta\sigma_t}\sqrt{B}
\end{displaymath}

{
\setlength{\parindent}{0cm}
$\Phi$ is the Cumulative Distribution Function (CDF) of the probability density function of $\Delta s_t$.
}

Therefore, we utilize the above equation to calculate the expected state value $E_B^{s_t}(s_{t+1})$ when updating with different batch size orders over N iterations. Suppose the orders $O_B$ and $O_D$ start at iteration A. The expected state value of $O_B$ is $E_B^{s_{A+N}}$, which needs to be calculated N times using Equation (1) from $s_A$. Meanwhile, the expected state value of $O_D$ is $E_{k_mB}^{s_{A+N}}$, which needs to be calculated m times using Equation (1) from $s_A$. Afterwards, we quantify the difference in convergence between the \verb|DeFT| scheduling scheme and the baseline through the ratio of $E_B^{s_{A+N}}$ and $E_{k_mB}^{s_{A+N}}$. Table \uppercase\expandafter{\romannumeral5} shows an example result of $E_B^{s_t}(s_{t+1})$ from iteration A to iteration A+N when using batch size orders $O_B$ and $O_D$ in ResNet-101 training. The ratio of $E_{O_B}$ and $E_{O_D}$ approximates 1, indicating that the convergence of $O_B$ and $O_D$ is almost the same. Note that although \verb|DeFT| showed a small loss compared to baseline in the results of Table \uppercase\expandafter{\romannumeral5}, it does not mean that the final training accuracy will suffer an equal amount of loss, as the convergence of the model fluctuates during training. The experiments in Section \uppercase\expandafter{\romannumeral5}.B show that our method achieves comparable training accuracy to the baseline, with negligible losses.

\subsubsection{Feedback mechanism}
The $Preserver$ module in \verb|DeFT| utilizes the aforementioned approach to quantify the loss from the scheduling results output by $Solver$. Based on this approach, we implement a feedback mechanism to preserve the accuracy of the scheduling scheme. As shown in Fig. 7, after the $Solver$ module outputs the scheduling result, \verb|DeFT| first collects the convergence information to calculate $E_B(s_{t+N})$ and $E_{k_mB}^{s_{A+N}}$, including batch size, learning rate, etc. The $Preserver$ module then computes the ratio of the two and determines if it is within a range around 1, denoted as [1-$\varepsilon$,1+$\varepsilon$]. If the ratio is within this range, it implies that the scheduling result passes the convergence test, hereupon \verb|DeFT| immediately applies it to training. Otherwise, \verb|DeFT| increases the knapsack capacity of the problem in $Solver$, allowing for more communication in each iteration. Since the update frequencies of \verb|DeFT|'s scheduling schemes can be approximated to coverage rates, this increase makes the update frequencies closer to the baselines, thereby preserving accuracy. The $Solver$ then regenerates the scheduling result with the new knapsack capacity and repeats this process (up to ten times to avoid excessive overhead) until the convergence difference falls below the threshold.

In our experiments, $\varepsilon$ is empirically set to 0.01. If $\varepsilon$ is too small, it may result in too many retries of \verb|DeFT|'s $Solver$, which may significantly increase its overhead. In addition, due to the increased communication time for each retry, the computation cannot cover communication, resulting in a decrease in algorithm performance. On the contrary, if $\varepsilon$ is too large, it may cause a decrease in training accuracy.

\section{Evaluation}

\begin{figure*}[t]
\setlength{\abovecaptionskip}{-0.2cm}
  \centering
  \includegraphics[width=\textwidth]{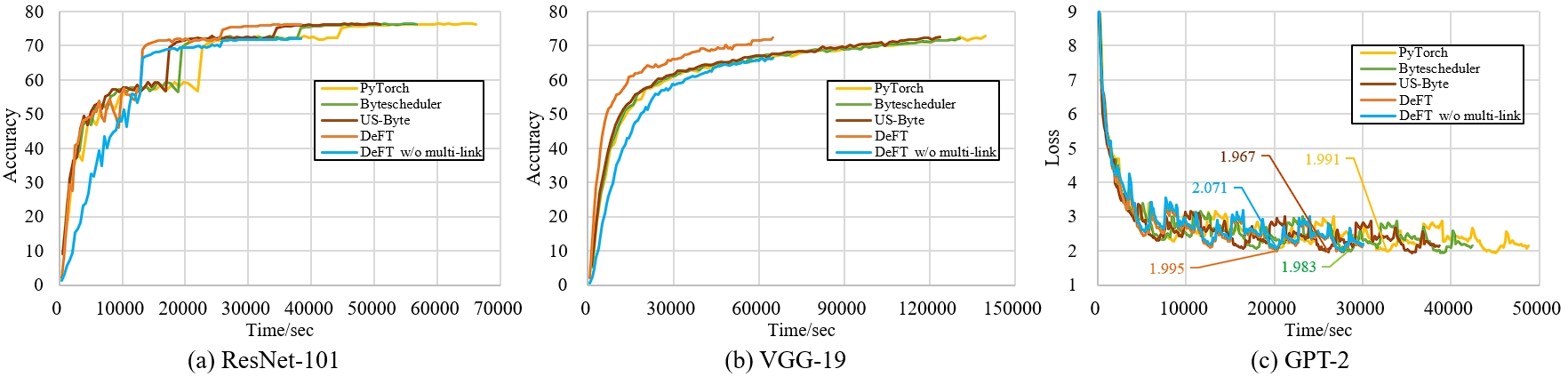}
  \caption{Time-to-solution curves of three DNNs on four schedulers with one ablation experiment.}
\end{figure*}

\begin{table}
\setlength{\abovecaptionskip}{0.cm}
\setlength{\belowcaptionskip}{-0.cm}
\setlength{\leftskip}{-2pt}
  \caption{Neural networks used for evaluation}
  \label{tab:evaluation}
  \resizebox{\linewidth}{8mm}{
  \begin{tabular}{cccc}
    \hline
    Task&DNN&parameters&datasets\\
    \hline
    \multirow{2}{*}{Image classification}&ResNet-101&44654504&\multirow{2}{*}{ImageNet}\\
    &VGG-19&143652544&\\
    Text generation&GPT-2&81894144&THUC-News\\
  \hline
\end{tabular}}
\end{table}

\subsection{Experimental setups}
{
\setlength{\parindent}{0cm}
$\mathbf{Testbed}$ $\mathbf{setup.}$ Our GPU cluster contains 2 nodes, each node contains 8 NVIDIA A100 GPUs with PCIe connections and equipped with two 40Gbps Ethernet network interface cards.
}

{
\setlength{\parindent}{0cm}
$\mathbf{Benchmarks.}$ We use three neural networks from different deep learning domains summarized in Table \uppercase\expandafter{\romannumeral6} for evaluation.
}

{
\setlength{\parindent}{0cm}
$\mathbf{Baselines.}$ We compare \verb|DeFT| with 3 representative baselines. (1) Pytorch DistributedDataParallel\cite{b18,b14}, which implements WFBP and tensor fusion with a default bucket size 25 MB; (2) Bytescheduler \cite{a1}, a priority scheduling strategy which overlaps communications with forward propagation. (3) US-Byte\cite{a6}, which proposes a greedy algorithm with low complexity to find an approximate optimal solution of scheduling order. 
}

\subsection{Time-to-Solution}
We collected time-to-solution curves for the four scheduling schemes during the training of three benchmarks in order to verify the convergence and performance of these schemes. The tensor partition size for Bytescheduler, US-Byte, and \verb|DeFT| was set to 6,500,000 in this subsection, allowing for a fair comparison with the default bucket size of 25 MB in PyTorch DDP.

\subsubsection{ResNet-101}
Fig. 10(a) presents the time-to-solution curves of ResNet-101. The $Multistep$ learning rate scheduler decays the learning rate by 10 times every 30 epochs. \verb|DeFT| outperformed the other three schemes by 33\%-73\% in speedup while achieving almost no loss in training accuracy. 

Fig. 11 presents schematic diagrams illustrating the scheduling difference of the four different scheduling schemes in ResNet-101 training. Because the computation and communication time for some buckets may be extremely short, we have adjusted their length in the figure for easy understanding. It can be observed that due to ResNet-101's coverage rate of approximately 1.4, the forward and backward computation times cannot fully cover all communication times. Even though both Bytescheduler and US-Byte almost utilize all forward and backward time to overlap communication, there are still significant bubbles in the forward computing. In contrast, \verb|DeFT| indirectly reduces the coverage ratio to around 1.0 by adopting heterogeneous communication, and then eliminates data dependencies to achieve fully overlapped scheduling. In the next iteration of the forward stage, due to the delayed update, the model is trained using the parameters from the previous iteration (marked with red letters).

\begin{figure}[t]
\setlength{\abovecaptionskip}{-0.1cm}
\setlength{\belowcaptionskip}{-0.cm}
  \centering
  \includegraphics[width=\linewidth]{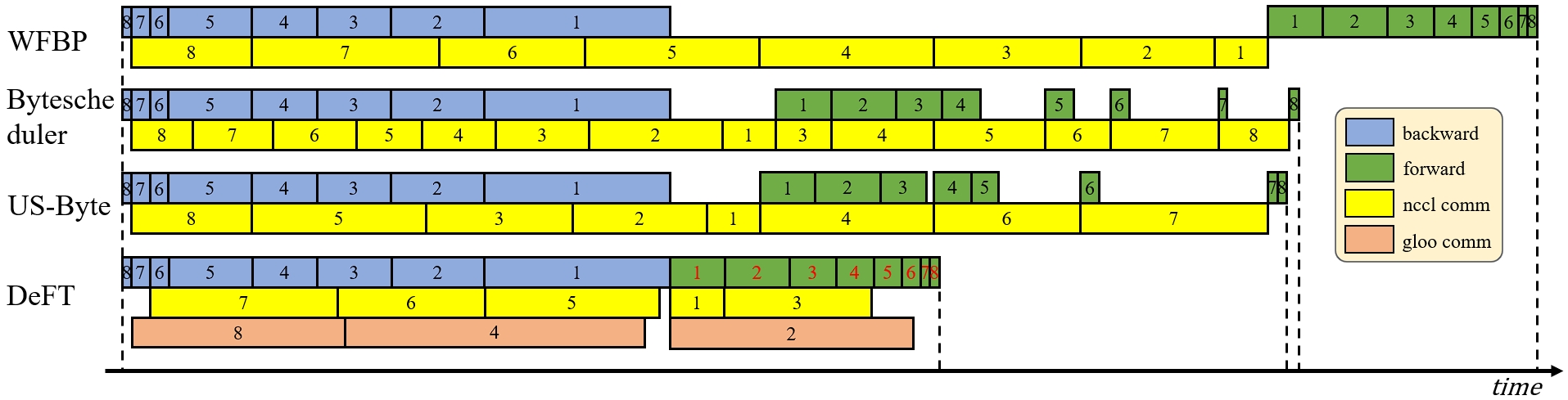}
  \caption{Bucket scheduling orders of different schemes in ResNet-101 training.}
\end{figure}

\begin{figure}[t]
\setlength{\abovecaptionskip}{-0.1cm}
\setlength{\belowcaptionskip}{-0.cm}
  \centering
  \includegraphics[width=\linewidth]{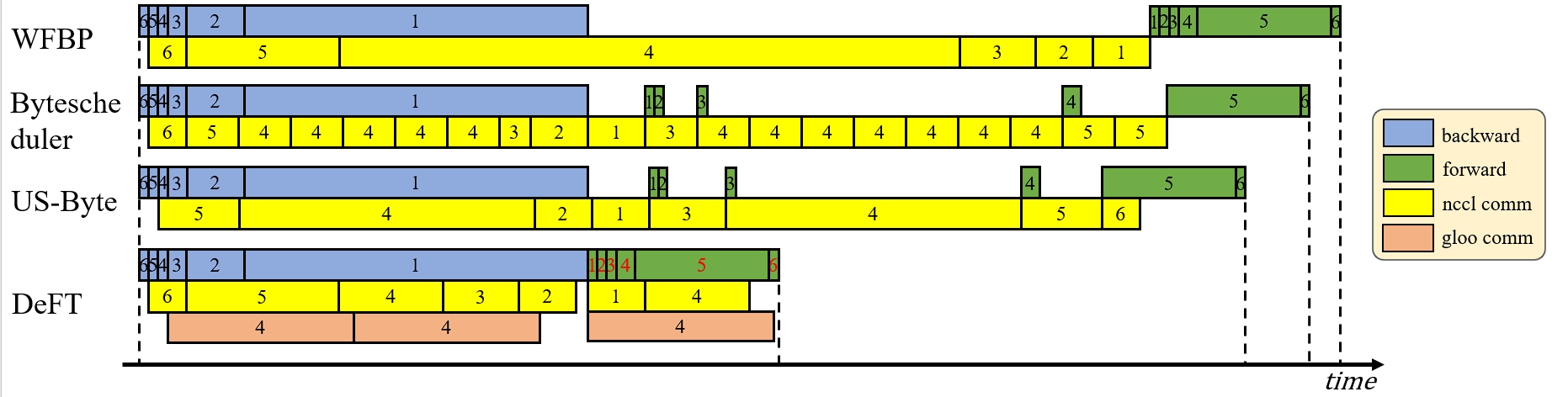}
  \caption{Bucket scheduling orders of different schemes in VGG-19 training.}
\end{figure}

\begin{figure}[t]
\setlength{\abovecaptionskip}{-0.1cm}
\setlength{\belowcaptionskip}{-0.cm}
  \centering
  \includegraphics[width=\linewidth]{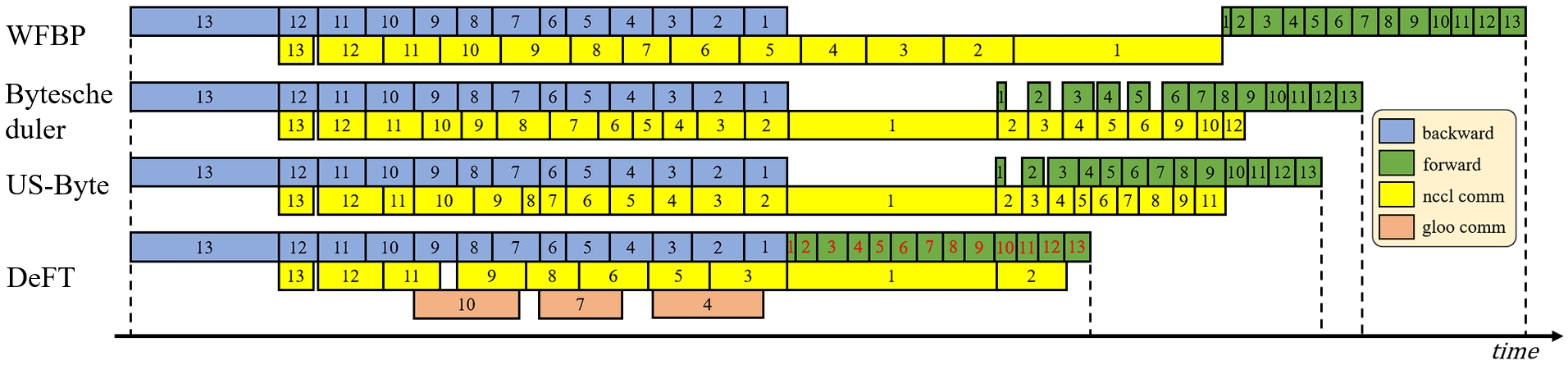}
  \caption{Bucket scheduling orders of different schemes in GPT-2 training.}
\end{figure}

\begin{figure*}[t]
\setlength{\abovecaptionskip}{-0.2cm}
\setlength{\belowcaptionskip}{-0.cm}
  \centering
  \includegraphics[width=\textwidth]{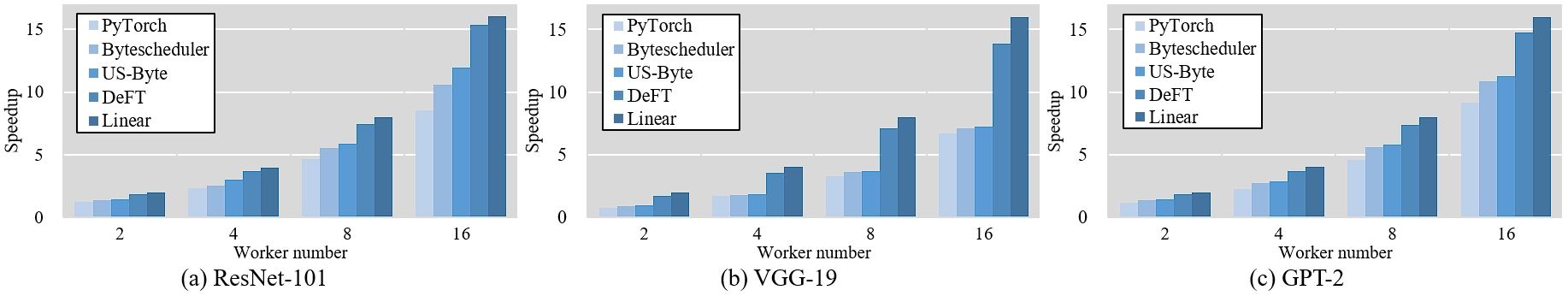}
  \caption{Speedups of different scheduling schemes under four different GPU numbers}
\end{figure*}

\subsubsection{VGG-19}
Fig. 10(b) presents the time-to-solution curves of VGG-19. In the experiments for VGG-19, \verb|DeFT| outperformed the other three schemes by 90\%-115\% in speedup. Compared to the experiments on ResNet-101, \verb|DeFT| accelerated more significantly on VGG-19 because VGG-19 has approximately 3.2 times the number of parameters as ResNet-101, with a coverage rate (CR) of approximately 2.0. Therefore, \verb|DeFT|'s scheduling scheme correspondingly lowers the update frequency to save more communication time.

As shown in Fig. 12, VGG-19 also suffers from high coverage rates and data dependency issues, resulting in significant bubbles. Additionally, the imbalance between bucket computation and communication prevents the full utilization of overlapping opportunities. For instance, while bucket \#5 has a longer forward computation time, the communication time for bucket \#6 is shorter. Since only the communication of bucket \#6 can overlap with the forward computation of bucket \#5, some overlapping opportunities are wasted. In contrast, \verb|DeFT| disregards the impact of data dependencies to flexibly utilize overlapping opportunities.

\subsubsection{GPT-2}
Fig. 10(c) presents the time-to-solution curves for GPT-2 and the metric uses training loss instead of accuracy. Although the parameter number of GPT-2 is about twice that of ResNet-101, its coverage rate is only about 0.99 because of the high computing overhead of transformer blocks. \verb|DeFT| outperforms other schemes by 29\%-62\%. 

Different from the previous two DNNs, GPT-2 has a coverage rate of approximately 1, meaning the total computation time is sufficient to overlap most of the communication overhead. Additionally, the computation and communication times of GPT-2's buckets are relatively balanced as shown in Fig. 13. However, due to hard dependency issues, the backward computation time of bucket \#13 and the communication time of bucket \#1 cannot be overlapped, which are relatively long, accounting for about 25\% of the iteration time. Due to the inability of Bytescheduler and US-Byte to take advantage of this overlapping opportunity, their performance cannot achieve optimum. In contrast, \verb|DeFT| delays the communication of bucket \#1 to the next iteration's forward stage, so that it can be fully overlapped.

\subsubsection{DeFT without heterogeneous multi-link communication}
We conducted an ablation study on the heterogeneous multi-link communication in Section \uppercase\expandafter{\romannumeral3}.C. The purpose of multi-link is to reduce the communication ratio (CR) and prevent accuracy degradation caused by excessive reduction of the DeFT parameter update frequency. To test the training accuracy without using multiple links (i.e., with a higher CR), we temporarily disabled the accuracy preservation strategy in $Preserver$. As shown in Fig. 10, although DeFT still maintained a relatively fast training speed (similar to the multi-link scenario) by adaptively reducing the update frequency, the training accuracy decreased on all three models. Specifically, the final accuracy of ResNet-101 decreased from 76\% to 71\%, while that of VGG-19 decreased from 71\% to 66\%. For GPT-2, the final loss was similar to other approaches, but the convergence speed was significantly slower in the early stages, as marked in Fig. 10(c).

\subsection{Scalability}

We employed different numbers of GPUs for training to validate the scalability of \verb|DeFT|. The tensor partition sizes in US-Byte and Bytescheduler were also uniformly set to 6,500,000. We will discuss the impact of partition size on scheduling results in Section \uppercase\expandafter{\romannumeral5}.E. We calculated the relative speedup by dividing the throughput of training on multiple GPUs by it on a single GPU, and the results are shown in Fig. 14. "Linear" represents optimal performance of DP. It can be observed that \verb|DeFT| achieved the best performance among four scheduling schemes across the three DNNs under four different GPU numbers. Specifically, the speedup of \verb|DeFT| was 1.21-1.92x of US-Byte, 1.32-1.98x of Bytescheduler, and 1.55-2.24x of PyTorch across the three DNNs.

\begin{figure*}[t]
\setlength{\abovecaptionskip}{-0.2cm}
\setlength{\belowcaptionskip}{-0.cm}
  \centering
  \includegraphics[width=\textwidth]{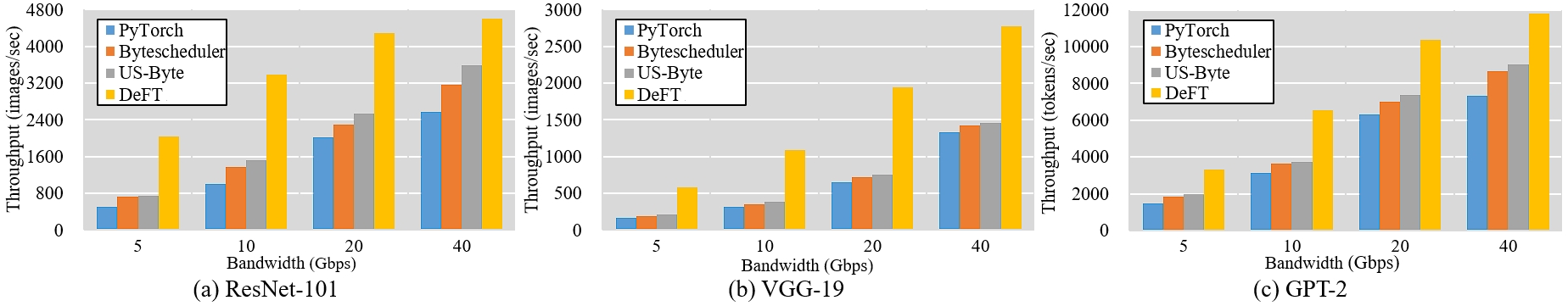}
  \caption{Throughput of different scheduling schemes under four different bandwidths}
\end{figure*}

\subsection{Performance under different bandwidths}
We studied the throughput of \verb|DeFT| at different bandwidths in this subsection. We used Linux tc tools to change the communication bandwidth between nodes. Due to the limitation of hardware resources, the maximum Ethernet connection bandwidth between nodes is 40 Gbps. The results are shown in Fig. 15. \verb|DeFT| achieved the highest acceleration compared to other solutions across all bandwidth levels, benefiting from its lower bubble ratio. Specifically, the speedup of \verb|DeFT| was 1.28-2.83 times of US-Byte, 1.36-3.09 times of Bytescheduler, and 1.61-3.94 times of PyTorch across four different bandwidths. Additionally, due to the $Preserver$ module restricting the reduction of the update frequency, when the bandwidth decreases, the speedup of \verb|DeFT| will be linearly related to the bandwidth, following the same trend as Bytescheduler and US-Byte.

\subsection{The influence of partition size}
We further analyzed the impact of partition size on the scheduling results. We selected VGG-19 as the test DNN due to its representative results. In addition to the scheduling results with a partition size set to 6,500,000, as described in Section \uppercase\expandafter{\romannumeral5}.B, we also analyzed the scheduling results for these schemes with four other partition sizes set to 3,000,000, 4,000,000, 8,000,000, and 10,000,000, respectively. We correspondingly adjusted the $bucket\_size\_mb$ parameter of PyTorch DDP to 10, 15, 30, and 40 MB for a fair comparison. Figure 16 illustrates the differences in the scheduling results in these four settings.

In Fig. 16(a) and Fig. 16(b), the partition and bucket sizes are relatively small, resulting in higher frequency of communication. Although Bytescheduler effectively utilizes forward time to overlap communication, excessively frequent communications significantly extend its overall communication time. In contrast, the US-Byte fusion strategy can significantly reduce total communication overhead when partition size is low.

\verb|DeFT| adopts the fusion strategy of US-Byte, but its total communication time is not the lowest among the four schemes, as the communication time of the fused tensor cannot exceed the forward time divided by $\mu$ (the speed ratio of the two heterogeneous communication links as described in Section \uppercase\expandafter{\romannumeral3}.C). However, \verb|DeFT| achieves the optimal performance among the four schemes through its strategies such as heterogeneous communication links, delayed updates, and dynamic update frequencies. For example, in Figure 16.c, a portion of bucket \#4 is delayed for communication in the subsequent iteration’s backward phase, and the new buckets in the next iteration (indicated in semi-transparent) are correspondingly delayed and stored locally. These buckets will be merged after the cumulative quantity includes all buckets of one iteration, as described in Section \uppercase\expandafter{\romannumeral3}.B.

\begin{figure*}[t]
\setlength{\abovecaptionskip}{-0.2cm}
\setlength{\belowcaptionskip}{-0.cm}
  \centering
  \includegraphics[width=\textwidth]{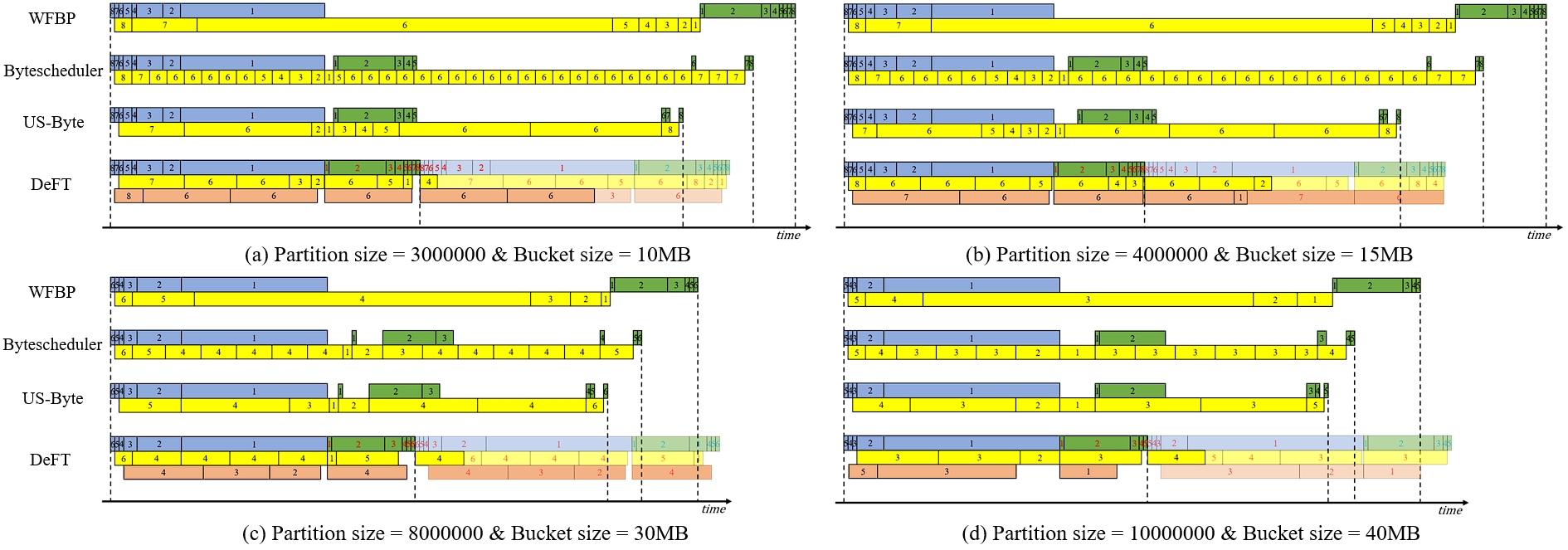}
  \caption{Scheduling results of four schemes with different partition size in VGG-19 training.}
\end{figure*}

\section{Limitations and Future Works}
\verb|DeFT| is a communication scheduling scheme that maximizes the overlap of communication and computation in data parallelism by mitigating data dependencies. Its advantages include:

\begin{itemize}[]
    \item \textbf{Overcoming Data Dependency Limitations:} \verb|DeFT| mitigates data dependencies in DP for more flexible scheduling.
    \item \textbf{Adaptive Adjustment of Update Frequencies:} \verb|DeFT| adaptively adjusts update frequencies based on the coverage rate to accommodate different tasks.
    \item \textbf{Better Trade-off:} \verb|DeFT| makes a better trade-off between performance and accuracy through its feedback mechanism.
\end{itemize}

Despite its strengths, \verb|DeFT| has limitations. Primarily, its performance still depends on the coverage rate of the tasks, which makes it unsuitable for certain scenarios. Communication scheduling schemes can alleviate communication bottlenecks in data parallelism but are ineffective for tasks with low coverage rates. For example, recent popular transformer models have long computation times. We attempted to train \textbf{Llama-2} 7B \cite{a27} with \verb|DeFT|. However, due to an extremely low coverage rate ($<$ 0.1), communication scheduling schemes such as \verb|DeFT| were unable to achieve improvement.

Furthermore, in scenarios with low bandwidth and extremely high coverage rates such as federated learning or edge computing, \verb|DeFT| proactively reduces update frequencies, which may lead to a decrease in accuracy. Due to the convergence feedback mechanism of the $Preserver$ module, \verb|DeFT| repeatedly expands the knapsack capacity and utilizes its $Solver$ to find the optimal scheduling. In low bandwidth scenarios, such approach could significantly increase the algorithmic overhead. Additionally, in environments with poor network conditions, where tensor communication times may far exceed the knapsack capacity, our tensor partitioning strategy could lead to many more tensors. Due to the fixed overhead of the communication operators, this significantly increases the total communication overhead.

For heterogeneous multi-link communication, it may incur additional overhead due to \verb|DeFT| offloads tensors to CPU. To mitigate this, we plan to explore techniques such as utilizing CPU-pinned memory to establish a memory mapping between CPU and GPU memory, thereby reducing the D2H/H2D copy overhead in future works.

\section{Related Work}
{
\setlength{\parindent}{0cm} $\mathbf{Communication}$ $\mathbf{Acceleration}$. Various software techniques have been proposed to address communication bottlenecks in distributed deep learning. For instance, many communication libraries such as MPI, gloo \cite{a14}, and NCCL \cite{a15} have implemented high-performance collective communication to support efficient communication between GPUs or other links. gZCCL \cite{a32} is a GPU-aware, compression-enabled collective communication library with accuracy-aware error control. SCCL \cite{a33} is a component in Microsoft Collective Communication Library (MSCCL), synthesizes collective communication algorithms tailored to the hardware topology. TACCL \cite{a34} is a tool that synthesizes efficient collective communication algorithms for specific hardware using a novel sketch abstraction. SYNDICATE \cite{a35} maximizes overlapping with computation in large-scale ML training by breaking down large communication tasks into smaller motifs.}

In addition, some work focuses on fine-grained optimization of collective communication operators, such as applying topology awareness \cite{a22,a23}, splitting operators \cite{a25}, and scheduling chunks \cite{a24}. Other approaches involve reducing data communication during synchronization processes through gradient compression techniques such as quantization \cite{b19,b11} and sparsification \cite{a26,b16,b13}. \verb|DeFT| is orthogonal to these approaches and can be integrated with them to further enhance communication acceleration.

{
\setlength{\parindent}{0cm}
$\mathbf{Communication}$ $\mathbf{Scheduling}$. The communication scheduling schemes partition, fuse, and reorder communications by exploiting the dependencies between computation and communication in data parallel training. Their goal is to maximize the overlapping between communication and computation. MG-WFBP \cite{b21} merges gradients based on the timing of backward propagation and communication for each layer; ASC-WFBP \cite{a12} enhances network bandwidth utilization by employing concurrent communication; P3 \cite{a2}, Bytescheduler \cite{a1}, and similar approaches utilize priority scheduling to exploit the overlap capacity of forward computation; PACE \cite{a4} and Prophet \cite{a5} discuss optimal tensor partition/fusion strategies within priority scheduling; US-Byte \cite{a6} demonstrates that the order of priority scheduling is not optimal and proposes a low-complexity greedy algorithm to find an approximately optimal schedule for forward propagation. In contrast to the aforementioned methods, \verb|DeFT| boldly loosens the dependencies in data parallelism to completely eliminate computational bubbles in communication scheduling schemes and devises a convergence guarantee strategy to achieve better performance.
}

\section{Conclusion}
In this paper, we present a communication scheduling strategy called \verb|DeFT|. \verb|DeFT| transforms the scheduling problem into a 0/1 knapsack problem, thereby scheduling all communication in a fully overlapping manner with computation. By mitigating the dependency constraints in data parallelism, \verb|DeFT| eliminates the bubbles caused by hard dependencies and reduces the coverage rate by adaptively lowering the update frequency. Furthermore, \verb|DeFT| introduces heterogeneous communication links for concurrent communication, thereby utilizing more communication resources and avoiding excessive reduction in update frequency that could lead to decreased accuracy. Additionally, \verb|DeFT| designs a $Preserver$ module to ensure the training accuracy of its scheduling scheme. Implemented within the PyTorch framework, \verb|DeFT| outperforms popular scheduling schemes in extensive experiments on clusters.

\section*{Acknowledgments}
{
\setlength{\parindent}{0.4cm}
This work is supported in part by Science and Technology Innovation 2030 - Major Project (No. 2022ZD0119104).
}




\end{document}